\documentclass[journal,10pt,letter]{IEEEtran}
\usepackage{amsmath,amssymb,amsfonts,mathrsfs,bm,amsthm}
\usepackage{amstext}
\usepackage{upgreek}

\usepackage{multicol}
\usepackage{indentfirst}
\usepackage{graphicx}
\usepackage{paralist}
\usepackage{multirow}
\usepackage{tabularx}
\usepackage[noadjust]{cite}

\usepackage{booktabs}
\usepackage{subfigure}
\usepackage{color}
\definecolor{blue}{rgb}{0,0.24,0.54}
\usepackage{textcomp}

\IEEEoverridecommandlockouts
\allowdisplaybreaks
\usepackage{algorithm,algorithmicx}
\floatname{algorithm}{\bf \small Algorithm}

\usepackage[noend]{algpseudocode}
\algrenewcommand{\algorithmiccomment}[1]{\hfill \# #1}

\newcommand\Algphase[1]{%
	\vspace*{-.7\baselineskip}\Statex\hspace*{\dimexpr-\algorithmicindent-2pt\relax}\rule{0.49\textwidth}{0.4pt}%
	\Statex\hspace*{-\algorithmicindent}\textbf{#1}%
	\vspace*{-.7\baselineskip}\Statex\hspace*{\dimexpr-\algorithmicindent-2pt\relax}\rule{0.49\textwidth}{0.4pt}%
}

\makeatletter
\newcommand{\algmargin}{\the\ALG@thistlm}
\makeatother
\algnewcommand{\parState}[1]{\State 
	\parbox[t]{\dimexpr\linewidth-\algmargin}{\strut #1\strut}}

\usepackage{accents}

\allowdisplaybreaks
\begin{document}
\title{\Large \vspace*{-5mm} \bf  Deep Reinforcement Learning Aided Packet-Routing For Aeronautical Ad-Hoc Networks Formed by Passenger Planes} %
\author{{Dong Liu, Jingjing Cui, Jiankang Zhang, Chenyang Yang, and Lajos Hanzo}
 	\vspace{-8mm}
	\thanks{D. Liu, J. Cui, and L. Hanzo are with the School of Electronics and Computer Science, the University of Southampton, Southampton SO17 1BJ, U.K. (e-mail: d.liu@soton.ac.uk; jingj.cui@soton.ac.uk; lh@ecs.soton.ac.uk).} 
	\thanks{J. Zhang is with the Department of Computing and Informatics, Bournemouth University, Bournemouth BH12 5BB, U.K. (e-mail: jzhang3@bournemouth.ac.uk).}
	\thanks{C. Yang is with the School of Electronics and Information Engineering, Beihang University, Beijing 100191, China (e-mail: cyyang@buaa.edu.cn).}
	\thanks{This work was supported in part by the Engineering and Physical Sciences
		Research Council Projects under Grant EP/N004558/1, Grant EP/P034284/1,
		Grant EP/P034284/1, and Grant EP/P003990/1 (COALESCE), in part by the Royal Society’s Global Challenges Research Fund Grant, and in part by the European Research Council’s Advanced Fellow Grant QuantCom (Grant No. 789028).}
	\thanks{The code for reproducing the results of this paper is available at https://github.com/Fluidy/tvtl2020.}
}

\maketitle

\vspace{-4mm}
\begin{abstract}
Data packet routing in aeronautical ad-hoc networks (AANETs) is challenging due to their high-dynamic topology. In this paper, we invoke deep reinforcement learning for routing in AANETs aiming at minimizing the end-to-end (E2E) delay. Specifically, a deep Q-network (DQN) is conceived for capturing the relationship between the optimal routing decision and the local geographic information observed by the forwarding node. The DQN is trained in an offline manner based on historical flight data and then stored by each airplane for assisting their routing decisions during flight. To boost the learning efficiency and the online adaptability of the proposed DQN-routing, we further exploit the knowledge concerning the system's dynamics by using a deep value network (DVN) conceived with a feedback mechanism. Our simulation results show that both DQN-routing and DVN-routing achieve lower E2E delay than the benchmark protocol, and DVN-routing performs similarly to the optimal routing that relies on perfect global information.  
\end{abstract}

\vspace{-1mm}

\vspace{-1mm}
\section{Introduction}
Next-generation wireless systems are expected to support global communications, anywhere and anytime~\cite{huang2019airplane}. Current in-flight Internet access supported by geostationary satellites or direct air-to-ground (A2G) communications typically exhibit either high latency or limited coverage. Aeronautical ad-hoc networks (AANETs) are potentially capable of extending the coverage of A2G networks by relying on commercial passenger airplanes to act as relays for forming a self-configured wireless network via multihop air-to-air (A2A) communication links~\cite{zhang2019aeronautical}.

Due to the high velocity of aircraft and the distributed nature of ad-hoc networking, one of the fundamental challenges in AANETs is to design an efficient  routing protocol for constructing an appropriate path for data transmission at any given time. 
Traditional topology-based ad-hoc routing protocols~\cite{li2007routing} usually require each node to locally store a routing table specifying the next hop. The routing table, however, has to be refreshed whenever the network topology changes during a communication session, hence imposing substantial signaling overhead and latency in AANETs. Although research efforts have been invested for improving the stability of routing in AANETs~\cite{sakhaee2006global,luo2017multiple}, they have a limited ability to update the routing tables for prompt adaption in high-dynamic scenarios. 

By contrast, another family of ad-hoc routing protocols, namely position-based (or geographic) routing~\cite{mauve2001survey}, only requires the position information of the single-hop neighbors and of the destination for determining the next hop. Since it does not have to maintain routing tables, geographic routing finds new routes almost instantly, when the topology changes. Because the  position information required can be readily obtained by each airplane using the automatic dependent surveillance-broadcast system on board, geographic routing is more appealing in AANETs. Greedy perimeter stateless routing (GPSR)~\cite{karp2000gpsr} was one of the most popular geographic routing protocols, which has also inspired various extensions~\cite{medina2011geographic,wang2013gr} in AANETs. The core idea of greedy routing is to forward the packet to the specific neighbor that is geographically closest to the destination. In \cite{medina2011geographic}, greedy routing was improved for avoiding congestion by considering the queue status of next hop. In \cite{wang2013gr}, the mobility information was further taken into consideration for choosing a more stable next hop. However, the performance of greedy routing~\cite{karp2000gpsr,medina2011geographic,wang2013gr} suffers when no neighbor is closer to the destination than the forwarding node (such a situation is term as the \emph{communication void}).

To elaborate, the limitation of greedy-based routing arises from the fact that the nodes are unaware of the entire network topology. Therefore, our ambitious goal is to enable the forwarding node to infer the global topology from its local observation for bypassing the communication void more efficiently.  Although the topology of AANETs revolves dynamically, it exhibits certain patterns, since the flight path and takeoff time are preplanned and remain fairly similar on the same day of different weeks. This suggests that the local geographic information may be strongly correlated with the whole topology, and such correlation may be learned from historical flight data. In this context, recent advances in  deep reinforcement learning (DRL)~\cite{mnih2015human} have demonstrated the powerful capability of deep neural networks (DNNs) for learning a direct mapping from the observation gleaned to the desired action to be taken.

Against this background, we invoke DRL for routing in AANETs aiming at minimizing the end-to-end (E2E) delay. Our major contributions can be summarized as follows:
\begin{enumerate}
	\item We propose a DRL-based routing algorithm using deep Q-network (DQN)~\cite{mnih2015human}, for directly mapping local geographic information to optimal routing decisions. Distinguished from routing algorithms based on tabular-based reinforcement learning (RL), such as Q-routing~\cite{boyan1994packet} and its variants~\cite{mammeri2019reinforcement}, which requires frequent information exchange for updating the Q-table online whenever the topology changes, our proposed DQN can be trained offline based on historical flight data to ``embed" the global network topology. During its flight, the forwarding node can infer the information required for deciding the next-hop by inputting its local observation into the DQN, without requiring any online update.
	\item To boost the learning efficiency, we further design another routing algorithm based on deep value-network (DVN) by exploiting the knowledge concerning the system's dynamics. Moreover, we introduce a feedback mechanism so that the forwarding node is able to plan one step ahead for enhancing the online adaptability.
	\item Our simulation results show that both DQN-routing and DVN-routing achieve lower E2E delay than GPSR. Furthermore, the performance gap  between DVN-routing and the optimal routing relying on perfect global information is marginal.  
\end{enumerate}

\vspace{-2mm}
\section{System Model}
Consider an AANET formed by passenger airplanes. Two nodes can establish direct communication link when they are above the radio horizon. The delay of the direct link from node~$i$ to node $j$ can be expressed as 
\begin{equation}
D_{\rm link}(i,j) = \frac{d(i,j)}{c} + \frac{S}{R(i,j)}, \label{eqn:Dlink}
\end{equation}
where the first and second terms are the propagation delay and transmission delay, respectively, $d(i,j)$ denotes the distance between nodes $i$ and $j$, $c$ is the speed of light, $S$ is the packet size, and $R(i,j)$ denotes the data rate of the link $i \to j$. We consider the decode-and-forward relaying protocol and let $D_{\rm que}(i)$ denote the queuing delay at node $i$.

Our goal is to find the optimal route $\mathcal{P} = (i_1, \cdots, i_T)$ minimizing the E2E packet delay between an source node $i_{\rm s}$ and destination $i_{\rm d}$, which can be formulated as
\begin{subequations} \label{eqn:routing}
\begin{align}
\!\!\min_{\mathcal{P}} ~~&   \sum\nolimits_{t=1}^{T-1} \left[D_{\rm que}(i_t) + D_{\rm link}(i_t, i_{t+1}) \right] \\
s.t. ~~& I(i_t, i_{t+1}) = 1, ~ \forall t=1,\cdots, T-1,
\end{align}
\end{subequations}
where $I(i_t, i_{t+1})=1$ if nodes $i_t$ and $i_{t+1}$ are above the radio horizon and $I(i_t, i_{t+1})=0$ otherwise, $i_1 = i_{\rm s}$ and $i_{T} = i_{\rm d}$.

Problem~\eqref{eqn:routing} can be solved by classic shortest path search algorithms, which requires the global information regarding the queuing delay of each node and the link delay between every two nodes. However, the node positions change rapidly due to the high velocity of airplanes. Consequently, this may impose substantial signaling overhead by keeping the required information  up-to-date for implementing the  algorithm. 

In the following, we assume that each node is only aware of its own position, the positions of the nodes within its direct communication range (i.e., its \emph{neighbors}) as well as the destination, and invoke DRL for finding the optimal route in an distributed manner. 

\vspace{-2mm}
\section{DRL for Geographic Routing}
In this section, we first recast the routing problem \eqref{eqn:routing} into the RL framework by designing the key elements of RL, and propose our DRL-based routing policies. 

\vspace{-2mm}
\subsection{RL framework}
In a RL problem, an agent learns from its interactions with the environment for achieving a desired goal \cite{sutton1998reinforcement}. At each \emph{time step} $t$, the agent observes the \emph{state} $\mathbf s_t$ of the environment and on that basis executes an \emph{action} $a_t$. Then, the agent receives a \emph{reward} $r_{t+1}$ from the environment and transits into a new state $\mathbf s_{t+1}$. The goal of the agent is to learn a mapping from $\mathbf s_t$ to $a_t$ (i.e., a policy $\pi$) for minimizing\footnote{In contrast to a standard RL problem defined to maximize the return, we consider minimizing the return because we aim at minimizing the E2E delay.} an expected \emph{return} $\mathbb E \big[\sum_{t=1}^{T-1} \gamma^{t-1} r_{t+1}\big]$,  which reflects the accumulated reward received by the agent during an \emph{episode} with $T$ time steps.

In the routing problem~\eqref{eqn:routing}, we specify that a time step starts when a node has received a packet and ends when the packet has been transmitted to the next node. Then, the whole episode begins when the packet is generated by the source and ends when the packet has been received by the destination or fail to reach the destination within $t_{\max}$ hops. Different from existing RL-based routing algorithms that train a distinct agent for each individual node, the agent in our framework moves along with the packet, and all the nodes share the agent's parameters, which improves the learning efficiency and the scalability.  

Let $i_t$ denote the node where the packet is located at the beginning of time step $t$, and let $\mathbf x(i) = (\mathsf{longitude}_i~[\text{E}^\circ], \mathsf{latitude}_i~[\text{N}^\circ], \mathsf{altitude}_i~[\text{km}])$ denote the position of node $i$. Then, the current position of the packet and of the destination can be denoted as $\mathbf x(i_t)$ and $\mathbf x(i_{\rm d})$, respectively. 
Let $\mathcal{N}_{i_t} \triangleq \{j | I(i_t, j) = 1 \}$ denote the neighbors of node $i_t$. To limit the dimension of action exploration, the neighbors are ranked by their distances to the destination in ascending order, and the next hop is selected only from neighbors ranked among the top $K$. They are termed as the \emph{candidates} and denoted by $\mathcal{C}_{i_t} \triangleq \{i_t^1, \cdots, i_t^K\}$, where $i_t^{k}$ represents the $k$th-ranking neighbor (candidate) of node $i_t$.

{\bf Action}: In time step $t$, the forwarding node $i_t$ should determine which candidate is selected as the next hop. Therefore, the action $a_t$ can be represented by the ranking of the candidate to be selected. Then, the next hop is node $i_t^{a_t}$.

{\bf State}: Since our goal is to learn a routing policy that only depends on local geographical information. The state is designed to include the positions of the source and the destination, as well as on the positions of the candidates, i.e.,
\begin{equation}
\mathbf s_t = \left[\mathbf{x} (i_t) , \mathbf{x} (i_t^1), \cdots, \mathbf{x}(i_t^K), \mathbf x(i_{\rm d})\right] \triangleq \mathbf s(i_t), \label{eqn:state}
\end{equation}
where we introduce the notation $\mathbf s (i_t)$ to emphasize that $\mathbf s_t$ is the local geographic information observed by node $i_t$ and we will use $\mathbf s_t$ and $\mathbf s(i_t)$ interchangeably in the following.

{\bf Reward}: The reward function can be naturally designed as the delay experienced within time step $t$, 
\begin{equation}
r_{t+1} = D_{\rm que}(i_t) +D_{\rm link}(i_t,i_{t+1}). \label{eqn:r}
\end{equation}  
In this way, minimizing the return is equivalent to minimizing the average E2E delay. 

The \emph{action-value function} is defined as~\cite{sutton1998reinforcement}
\begin{equation}
Q_{\pi}(\mathbf s_t, a_t)  \triangleq \mathbb{E}\Big[ \sum\nolimits_{l=t}^{T-1} \gamma^{l-t} r_{l+1} ~\Big|~ \mathbf s_t, a_t, \pi \Big].
\end{equation}
For the  routing problem considered, we set $\gamma = 1$ and hence $Q_{\pi}(\mathbf s_t, a_t)$ represents the delay between the forwarding node and the destination by selecting node $i_t^{a_t}$ as the next hop and thereafter forwarding the packet according to policy~$\pi$. Then, the \emph{optimal action-value function} is defined as $Q_*(\mathbf s_t, a) \triangleq \min_{\pi} Q_{\pi} (\mathbf s_t, a_t)$, from which the optimal policy can be readily obtained as 
$\pi_*(\mathbf s_t) = \arg\min_{a} Q_*(\mathbf s_t, a)$. In this sense, $Q_*(\cdot)$ contains all the information required for determining the optimal next hop, or in other words, it embeds the global network topology. Therefore, the agent's goal can be accomplished by learning $Q_*(\cdot)$.
	
The Bellman equation for $Q_*(\cdot)$ can be expressed as~\cite{sutton1998reinforcement} 
\begin{equation}
Q_*(\mathbf s_t, a_t) = \mathbb{E} \big[ r_t + \min_{a} Q_*(\mathbf s_{t + 1}, a) ~\big|~ \mathbf s_t, a_t, \pi_*\big], \label{eqn:BQ}
\end{equation}
based on which various RL algorithms, such as Q-learning~\cite{sutton1998reinforcement}, have been developed to learn the optimal action value function.
However, Q-learning is faced with the curse of dimensionality due to the continuous nature of the state $\mathbf s_t$. Thus, we resort to DRL, specifically DQN, for learning the optimal action value function.

\vspace{-2mm}
\subsection{DQN-Routing}
We employ a DNN $Q(\mathbf{s}_t, a_t; \bm \theta_Q)$ shared by all the nodes to learn the optimal action-value function $Q_*(\mathbf s_t, a_t)$. The DQN parameter $\bm \theta_Q$ is trained offline using the historical flight trajectories. Specifically, we create a large set of snapshots containing the position of each flight at each timestamp. 

During the offline training phase, the transmission delay and propagation delay can be calculated based on the flight positions. As for the queuing delay, since we aim to train the DQN for embedding the historical topology information, which is independent from the packet traffic, we assume that the queuing delay is identical and constant among all the nodes during training. In this way, the total queuing delay is actually determined by the number of hops in the route.

Let each node forward its received packet in a $\varepsilon$-greedy manner, i.e., with probability $\varepsilon$ randomly selecting an action for exploration and with probability $1-\varepsilon$ selecting action $a_t = \arg\min_a Q(\mathbf s_t, a; \bm \theta_Q)$ for exploitation. Everytime a packet is forwarded to the next hop, the experience vector $\mathbf e_t = [\mathbf s_t, a_t, r_{t+1}, \mathbf s_{t+1}]$ is recorded in a replay memory $\mathcal{D}$ and we randomly sample a batch of experiences $\mathcal{B}$ from $\mathcal{D}$ for updating the parameter $\bm \theta_Q$ (i.e., the experience replay~\cite{mnih2015human}). Based on the Bellman equation~\eqref{eqn:BQ}, $\bm \theta_Q$ is updated by minimizing the loss function $\mathbb {E} [(y_t - Q(\mathbf s_t, a_t;\theta_Q))^2]$ using stochastic gradient descent
$
\bm \theta_Q \leftarrow \bm \theta_Q  - \frac{\delta}{|\mathcal{B}|} \nabla_{\bm \theta_Q} \sum\nolimits_{\mathbf e_l\in \mathcal{B}}\left[ y_l - Q(\mathbf s_{l}, a_l; \bm \theta_Q)\right]^2
$, where $\delta$ is the learning rate, $y_l  =  r_{l + 1} $  if the episode ends  
on state $\mathbf s_{l+1}$, and $y_l = r_{l+1} +  Q'(\mathbf{s}_{l+1}, \arg \min_{a}$ $   Q(\mathbf{s}_{l+1}, a; \bm \theta_Q); \bm \theta_Q')$ otherwise. Furthermore,~$Q'(\cdot; \bm \theta_Q')$ represents the target network, which has the same structure as the DQN $Q(\cdot; \bm \theta_Q)$ and is updated by $\bm \theta_Q' \leftarrow \tau \bm \theta_Q + (1-\tau) \bm \theta_Q' $ with very small value of $\tau$ to reduce the correlations between the action value $Q(\mathbf s_l, a_l;\bm \theta_Q)$ and the target values $y_l$~\cite{mnih2015human}. 

After the training converges, the DQN can be copied to each airplane in support of online routing decisions. During its flight, each airplane forwards its received packet according to the DQN based on the state it observes. Specifically, node $i_t$ observe its state $\mathbf s(i_t)$ and then evaluates 
\begin{equation}
a_t^* = \arg\min_{a \in \mathcal{A}_t} \left[Q(\mathbf s(i_t), a; \bm \theta_Q)\right], \label{eqn:atQ}
\end{equation}
where $\mathcal{A}_t \triangleq \{k| 1\leq k\leq K, i_t^k\neq i_1, \cdots, i_{t-1}\}$ specifics that the next hop cannot be chosen from the previously selected nodes to avoid loops in the routes. Then, node $i_t$ forwards its received packet to node $i^{a_t^*}$.

The above implementation of DQN represents a generic approach to solving completely model-free RL problems. However, in the considered routing problem, the system's dynamics can be partially known, which can be exploited for faster learning and better online adaptability. Moreover, the training of DQN treats the queuing delay as an identical constant, while in reality the queuing delay varies due to different packet arrival rate. In the following, we develop a specialized DRL algorithm for learning the optimal routing policy more efficiently and introduce a feedback mechanism for taking the real-time queuing delay into consideration.  

\vspace{-2mm}
\subsection{DVN-Routing With Feedback}

In this subsection, we first specify the knowledge concerning the system's dynamics, which is then exploited for boosting the learning efficiency. Then, based on the feedback received from the next-hop candidates, the forwarding node is capable of planning one step ahead before forwarding the packet, which improves the online adaptability of the policy.
\subsubsection{Exploiting the System's Dynamics}
For the routing problem of AANETs, given the current state $\mathbf s_t$ and an arbitrary action $a_t$, the next state can be predicted before the forwarding node sends the packet, because the movement of nodes can be neglected within a single time step.\footnote{For a flight cruise speed of $900$ km/h, the position shift within a typical time step of $10$ ms is only $2.5$ m, which is much smaller than the minimum distance allowed between airplanes and hence can be safely neglected.}
Specifically, the next state $\mathbf{s}_{t+1}$ is actually the state observed by node $i_t^{a_t}$ in time step $t$, which yields
\begin{equation}
\mathbf s_{t+1} = \mathbf s(i_t^{a_t}) = \big[\mathbf x(i_t^{a_t}), \mathbf x(i_t^{a_t,1}), \cdots, \mathbf x (i_t^{a_t, K}), \mathbf x(i_{\rm d}) \big], \!\! \label{eqn:st+1}
\end{equation}
where $i_{t}^{a_t, k}$ denotes the $k$th candidate of node $i_t^{a_t}$. 

As for the reward, before $i_t$ forwards a packet, the link delay $D_{\rm link}(i_t, i_t^{a_t})$ can actually be computed in advance according to~\eqref{eqn:Dlink} at the next hop $i_t^{a_t}$ and the queuing delay $D_{\rm que}(i_t^a)$ can also be measured by $i_t^a$ based on its queuing status~\cite{medina2011geographic}.

To exploit the above knowledge regarding the state transition and the reward, we introduce the \emph{intermediate-state-value function} of a  routing policy $\pi$, defined by
\begin{equation}
V_{\pi}(\mathbf s_t) \triangleq \mathbb{E}\Big[ D_{\rm link}(i_t, i_t^{a_t} ) + \sum\nolimits_{l=t+1}^{T} r_{l+1} ~\Big|~ \mathbf s_t, \pi \Big],
\end{equation}
which captures the expected delay commencing from the instant when the packet has experienced its queuing delay at node~$i_t$ until it reaches its final destination, by forwarding according to $\pi$. Correspondingly, the \emph{optimal intermediate-state-value function} is defined as $V_*(\mathbf s_t) \triangleq \min_{\pi} V_{\pi} (\mathbf s_t)$. 

Bearing in mind the definitions of $V_{*}(\cdot)$ and $Q_{*}(\cdot)$  as well as $r_{t+1}$, we can write $V_{*}(\cdot)$ in terms of $Q_{*}(\cdot)$ as 
\begin{equation}
V_*(\mathbf{s}_{t}) + \mathbb E\left[D_{\rm que}(i_t)\right]  = \min_{a}   Q_*(\mathbf s_t, a),   \label{eqn:VQ}\\
\end{equation}
and write $Q_{*}(\cdot)$ in terms of $V_{*}(\cdot)$ as
\begin{equation}
Q_*(\mathbf s_t, a)  =  \mathbb{E}\left[r_{t+1} + D_{\rm que}(i_t^{a})\right] + V_*(\mathbf{s}_{t+1}). \label{eqn:QV}
\end{equation}

Observe from \eqref{eqn:QV} that the value of $Q_*(\cdot)$ can be obtained by learning $V_*(\cdot)$ instead. Then, by substituting \eqref{eqn:QV} into \eqref{eqn:VQ} and considering $\mathbf s_{t+1} = \mathbf s(i_t^a)$, we can obtain the Bellman equation for $V_*(\cdot)$ as
\begin{equation}
V_*(\mathbf s_t) = \min_{a}\left\{ \mathbb{E}\left[{r} (i_t, i_t^a)\right] +   V_*\big(\mathbf s(i_t^a)\big) \right\}, \label{eqn:BV} 
\end{equation}
where $r  (i_t, i_t^a) \triangleq D_{\rm link}(i_t,i_t^{a})  + D_{\rm que}(i_t^{a})$.

\subsubsection{Offline Training}
Similarly to DQN-routing, we invoke a DNN $V(\mathbf s_t; \bm \theta_V)$ to learn $V_*(\mathbf s_t)$, termed as the DVN. In contrast to DQN, the scale of DVN can be much smaller, because it does not depend on the action, and hence has less parameters to train.  

During the offline training phase, again, we use the historical flight data and assume constant and identical queuing delay. Let each node forward its received packet in a $\varepsilon$-greedy manner. According to~\eqref{eqn:BV}, the action for exploitation is determined by
\begin{equation}
a_t =  \arg\min_a \big[ r(i_t, i_t^{a}) + V\big(\mathbf s (i_t^{a}); \bm \theta_V\big) \big]. \label{eqn:atV}
\end{equation}

Then, the experience vector composed by 
\begin{equation}
\tilde{\mathbf e}_t = [\mathbf s_t, \mathbf s(i_t^1), \cdots, \mathbf s(i_t^K),  r(i_t, i_t^1), \cdots,  r(i_t, i_t^K)] \label{eqn:xp}
\end{equation}
is recorded in the replay memory $\mathcal{D}$ and we randomly sample a batch of experiences $\mathcal{B}$ from $\mathcal{D}$. Based on the Bellman equation~\eqref{eqn:BV}, $\theta_V$ is updated by minimizing the loss function $\mathbb{E}[y_{t} - V(\mathbf s_{t}; \bm \theta_V)] $ via stochastic gradient descent as
\begin{equation}
\bm \theta_V \leftarrow  \bm \theta_V  + \frac{\delta}{|\mathcal{B}|} \sum\nolimits_{\tilde{\mathbf e}_l\in \mathcal{B}}\left[ y_{l+1} - V(\mathbf{s}_l; \bm \theta_V )\right]^2, \label{eqn:update}
\end{equation}
where $y_{l} = r(i_l, i_l^{a_*})$ if $i_l^{a_*} = i_{\rm d}$, $y_l = r(i_l, i_l^{a_*}) +  V'\big( \mathbf s (i_l^{a_*}); \bm \theta_V'\big)$ otherwise, $a_* = \arg\min_a \big[r(i_l, i_t^{a}) +  V\big(\mathbf s (i_l^{a}); \bm \theta_V\big) \big]$, and finally $V'(\cdot; \bm \theta_V')$ is the target network updated by $\bm \theta_V' \leftarrow \tau \bm \theta_V + (1-\tau) \bm \theta_V'$.

\subsubsection{Online Decision}
\begin{figure}[!htb]
	\vspace{-2mm}
	\centering
	\includegraphics[width=0.49\textwidth]{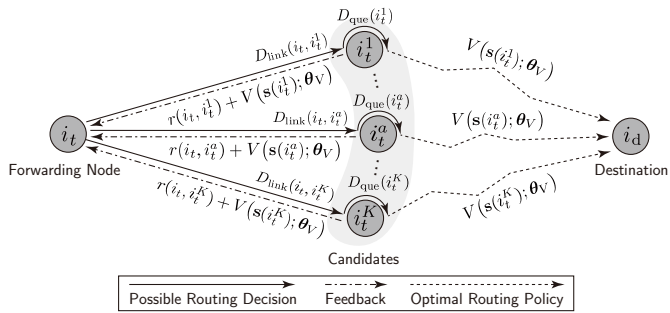}
	\vspace{-6mm}
	\caption{Illustration of the feedback mechanism in DVN-routing.}
	\label{fig:online}
\end{figure}

Once sufficiently well trained, the DVN is copied to each airplane for online routing decision. Since the information required for determining the action in \eqref{eqn:atV}, i.e., $r(i_t, i_t^{a})$ and $V\big(\mathbf s (i_t^{a}); \bm \theta_V\big)$ for $a=1, \cdots, K$, are only available at the next-hop candidates, we introduce a feedback mechanism for enabling the forwarding node $i_t$ to obtain these information. Specifically, each candidate $i_t^a$ estimates $D_{\rm link}(i_t, i_t^a)$ and $D_{\rm que}(i_t^a)$, observes its state $\mathbf s(i_t^a)$ and computes $V\big(\mathbf s(i_t^a); \bm \theta_V\big)$, and then sends $r(i_t, i_t^a) + V\big(\mathbf s(i_t^a); \bm \theta_V\big)$ to the forwarding node $i_t$, as shown in Fig.~\ref{fig:online}. Finally, the forwarding node $i_t$ selects the action
\begin{equation}
a_t^* = \arg\min_{a \in \mathcal{A}_t } \left[ r(i_t, i_t^a) +  V\big(\mathbf s(i_t^a); \bm \theta_V\big)\right], \label{eqn:at*}
\end{equation}
where $\mathcal{A}_t$ is used for avoiding loops in the routes.

Compared with directly determining the action based on the DQN by~\eqref{eqn:atQ}, the information used for deciding the action is observed by every next-hop candidate instead of that observed by the forwarding node alone. In this way, the forwarding node is able to plan one step ahead for more prompt adaption to the dynamic environment. For example, when the next-hop candidate $i_t^a$ has a higher traffic load, the queuing delay $D_{\rm que}(i_t^a)$ will increase, which increases the value of $r(i_t, i_t^a)$ and hence $i_t^a$ is less likely to be chosen as the next hop according to \eqref{eqn:at*}.

The whole learning and decision procedure is shown in Algorithm 1.

\begin{algorithm}[!htb] 
	\caption{DVN-Routing for AANETs}\
	\label{alg1}
	\small
	\begin{algorithmic}[1]
		\State Initialize $\bm \theta_V$ and $\bm \theta_V' \leftarrow \bm \theta_V$.
		\vspace{1mm}
		\Algphase{Offline DVN Training}\vspace{1mm}
		\For{${\tt episode} = 1, 2, \cdots, N$}
		\parState{Randomly sample a topology snapshot from historical flight data.}
		\State Set the source $i_{\rm s}$ and destination~$i_{\rm d}$.
		\For{$t = 1, 2, \cdots $}
		\If{$i_t^{a_t} = i_{\rm d}, t > t_{\max}$}
		\State {\bf break}
		\EndIf
		\parState{Observe state $\mathbf s_t = \mathbf s(i_{\rm s})$.}
		\parState{Randomly select action $a_t \in \{1,\cdots, K\}$ (with probability $\varepsilon$), or set $a_t = \arg\min_a\left[ r(i_t,i_t^{a}) +   V\big(\mathbf s(i_t^a); \bm\theta_V\big) \right]$ otherwise. }		
		\parState{Store the experience $\tilde{\mathbf e}_t$ composed by \eqref{eqn:xp} into $\mathcal D$.}
		\parState{Randomly sample a batch of experiences from~$\mathcal{D}$ as $\mathcal{B}$.}
		\State Update $\bm \theta_V$ and $\bm \theta_V'$ according to \eqref{eqn:update}.
		\EndFor
		\EndFor
		\vspace{1mm}
		\Algphase{Online Routing Decision}\vspace{1mm}	
		\Require $i_{\rm s}$, $i_{\rm d}$, $\bm \theta_V$.
		\For{$t = 1, 2, \cdots$}
		\If{$i_t = i_{\rm d}$}
		\State{\bf break}
		\EndIf
		\parState{The forwarding node $i_t$ observes $\mathbf s (i_t)$.}
		\For{$a=1,\cdots, K$}
		\parState{Node $i_t^{a}$ observes $\mathbf s(i_t^{a})$, estimates $r(i_t, i_t^a)$,  computes $V(\mathbf s(i_t^{a});\bm \theta_{V}) + r(i_t, i_t^a)$ and sends the result to node $i_t$.}
		\EndFor
		\parState{Node $i_t$ computes $a_t^*$ by~\eqref{eqn:at*} and forwards the packet to~$i_t^{a_t^*}$.}
		\EndFor
	\end{algorithmic}
\end{algorithm}

\section{Simulation Results}

In this section, we introduce the simulation environment and compare the performance of the  routing policies learned by DRL to benchmark policies via simulations. 

\vspace{-3mm}
\subsection{Simulation Environment}
\begin{figure}
	\vspace{-2mm}
	\centering	
	\subfigure[Flight positions at time $t_1$]{
		\label{fig:t1} 
		\includegraphics[width=0.35\textwidth]{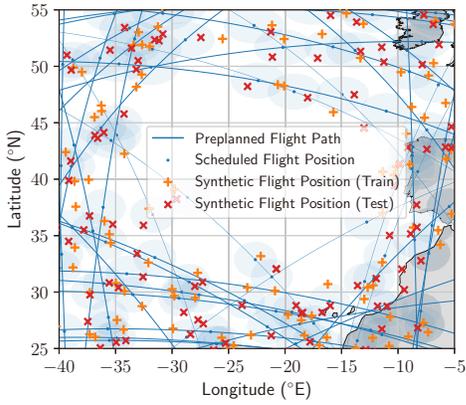} }
	\subfigure[Flight positions at time $t_{50}$]{
		\label{fig:t50} 
		\includegraphics[width=0.35\textwidth]{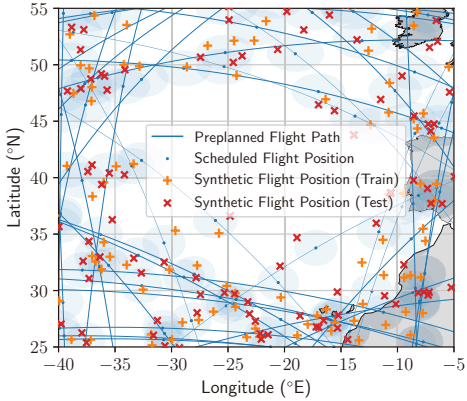}}\vspace{-2mm}
	\caption{Illustration of two example snapshots in training and testing set at different time. The shaded area represents the specific zone, where a particular flight may be found with probability $80\%$.}
	\label{fig:t}
\end{figure}

Since there is insufficient real flight data available for training and testing, we generate synthetic flight data in our simulation for mimicking the airplane mobility. 
Specifically, we consider a 3D-airspace within longitude $-40^\circ \sim -5^\circ$ East,  latitude $25^\circ \sim 55^\circ$ North, and altitude $0 \sim 13$ km, whose 2D-projection is shown in Fig.~\ref{fig:t}.
To reflect the typical non-uniform flight density distribution and to evaluate the performance of routing algorithms when communication void exists, we earmark a pair of no-fly zones located at $(-17.25^\circ, 40^\circ, 0)$ and $(-27.75^\circ, 44.5^\circ, 0)$ each having a radius of $500$ km and a height of $13$ km, where no flight path passes through. There are 40 preplanned great-circle flight paths randomly drawn through the available area and then fixed throughout the simulation to represent the seasonal flight corridors. A total of $100$ airplanes are uniformly placed along the 40 flight paths, where the airplanes on the same path are flying in the same direction with constant speed to maintain the safety flight separation distance. The altitude of each airplane is randomly chosen within the normal cruise altitude of $9 \sim 13$ km.

In reality, each airplane may not takeoff on time and may not fly strictly according its preplanned path due to varies reasons, which results in the mismatch between the historical flight positions (i.e., training data) and current flight positions (i.e., testing data). To reflect this issue, we add a random deviation to the \emph{scheduled flight position} (i.e., the position of airplane when it fly strictly according to the plan) to generate the \emph{synthetic flight position} for training and testing the routing algorithm, as shown in Fig.~\ref{fig:t}. Specifically, the random deviations along the latitude and longitude follow Gaussian distribution with a standard deviation of $100$ km.

We generate $2000$ snapshots of the network, half of which are used for training and the other half are used for testing outside the training set. In Fig.~\ref{fig:t}, we demonstrate a pair of example snapshots at different time. We can see that the flight positions change over time, and the  positions of the same flight in training and testing set are different.

In each snapshot, the source node is randomly selected, while the destination is set as the ground station is located at $(-10^\circ, 52^\circ, 0.05)$. The queuing delay is set as $D_{\rm que} = 5$~ms throughout the training. The packet size is $S = 15$ KB and the transmission data rate is configured according to the distance-based adaptive coding and modulation scheme of~\cite[Table~I]{zhang2017adaptive} using matched filter based beamforming relying on 32 transmit antennas and four receive antennas. 

\vspace{-2mm}
\subsection{Fine-Tuned Parameters of DQN- and DVN-Routing}
The DNNs are tuned as follows to achieve their best performance. The candidate set size is $K=10$. Both the DQN and DVN have two hidden layers, where each layer has $100$ and $50$ nodes for DQN and DVN, respectively. In this setting, the total number of unknown parameters to be learned in DVN is roughly reduced roughly by a factor of three compared to DQN. The hidden layers employ the rectified linear units (ReLU) as the activation function while the output layer has no activation function. In the training phase, the exploration probability is set as $\varepsilon = 1$ for the first $100$ episodes, decreases to $0.1$ within the next $400$ episodes, and remains $0.1$ for the rest of the episodes. The learning rate $\delta$ is $10^{-4}$ for both the DQN and DVN, while the update rate is $\tau = 10^{-3}$ for both the target networks. The batch size is $|\mathcal B| = 32$. During the testing phase, we set $\varepsilon = 0$ for both DQN and DVN, and the parameters $\bm \theta_V$, $\bm \theta_Q$ are frozen.

\vspace{-2mm}
\subsection{Performance Comparison}
The following benchmarks are considered for comparison:
\begin{itemize}
	\item Optimal: The optimal route found by solving problem~\eqref{eqn:routing} via the Floyd-Warshall algorithm, which relies on the global information regarding the link delay between every two nodes and the queuing delay of every single node.
	\item GPSR: The routing protocol proposed in~\cite{karp2000gpsr}, which is solely based on local geographic information. Specifically, each node forwards its received packet to the specific neighbor that is geographically closest to the destination. When a packet reaches a node where greedy forwarding fails, the algorithm recovers by routing around the perimeter of the region.
\end{itemize}

\begin{figure}
	\vspace{-2mm}
	\centering
	\includegraphics[width=0.35\textwidth]{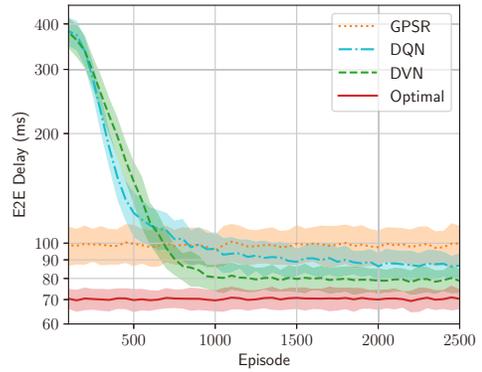}
	\vspace{-2mm}
	\caption{Learning curve in training phase. All experiments are run for 100 different random seeds each. The curves are smoothed by averaging over a window of 40 episodes. The lines reflect the average value and the shaded bands reflect the standard deviation.}
	\label{fig:train}
\end{figure}
In Fig.~\ref{fig:train}, we compare the learning curves of the proposed DQN-routing and DVN-routing algorithms during training. We can see that both the algorithms achieve lower average E2E delay than GPSR after $500$ episodes of training and finally approach the delay of optimal routing. Furthermore, since DVN-routing exploits the knowledge concerning the system's dynamics, it achieves lower E2E delay than DQN-routing after its convergence.

\begin{figure}
	\centering
	\includegraphics[width=0.35\textwidth]{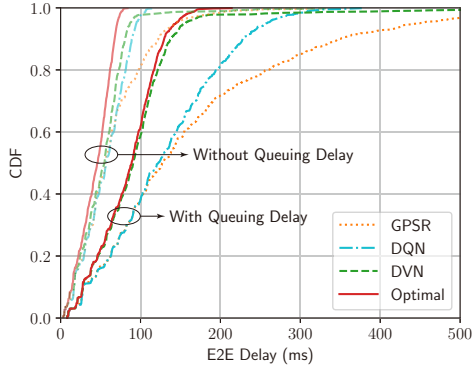}
	\vspace{-2mm}
	\caption{The CDF of E2E delay with or without considering the queuing delay.}
	\label{fig:cdf}
	\vspace{-2mm}
\end{figure}

In the online testing phase, the snapshots are generated outside the training set as previously mentioned to reflect the uncertainty in flight positions. Furthermore, to reflect the fluctuation of traffic load, $20\%$ of the nodes are randomly chosen to set with a higher queuing delay of $50$~ms. In Fig.~\ref{fig:cdf}, we compare the cumulative distribution function (CDF) curves of the E2E delay during the testing phase. We can see that upon neglecting the queuing delay in the E2E delay calculation, DQN achieves near-optimal performance. However, when taking the queuing delay into consideration, the gap between the optimal routing policy and DQN-routing increases. Nevertheless, DQN-routing still outperforms GPSR. By contrast, DVN-routing can still achieve near-optimal performance, even though it is trained offline assuming constant and identical queuing delay, because the feedback mechanism allows each next-hop candidate to report its real-time queuing delay to the forwarding node during the online decision phase.

\begin{figure}
	\centering
	\includegraphics[width=0.35\textwidth]{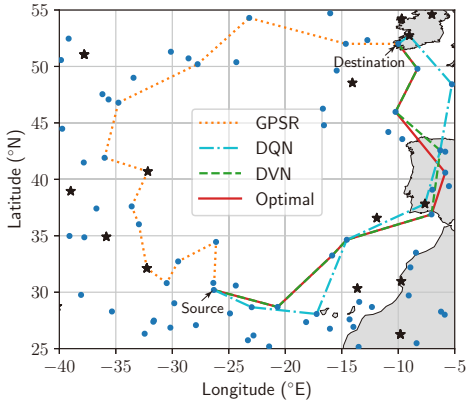}
	\vspace{-2mm}
	\caption{The routes found by each routing algorithms in a example snapshot, where ``$\bigstar$" denotes the congested nodes.}
	\label{fig:path}
\end{figure}
In Fig.~\ref{fig:path}, we show an example snapshot during the testing phase for comparing the routes found by different routing algorithms. We can see that GPSR is rather ``shortsighted" and struggles to get round the no-fly zone. By contrast, both DQN-routing and DVN-routing can find routes having a similar number of hops as the optimal routing policy, because they  implicitly exploit the network topology information that has been embedded in the DQN/DVN trained using historical flight trajectories. Since DQN-routing is unaware of the real-time queuing delay, it may encounter some congested nodes (marked by ``$\bigstar$") along the route. By arranging for each next-hop candidate to feed back its queuing delay  to the forwarding node, DVN-routing can find a route bypassing the congested relaying nodes.

\vspace{-2mm}
\section{Conclusions and Future Directions}
In this paper, we proposed DRL-based routing policies for minimizing the E2E delay in AANETs. We first used DQN for learning a direct mapping from the local geographic information to the optimal routing decision. To boost the learning efficiency and the online adaptability of the proposed DQN-routing, we additionally proposed DVN-routing by exploiting the knowledge concerning the system's dynamics and by introducing a feedback mechanism. Simulation results show that both DQN-routing and DVN-routing achieve lower E2E delay than  GPSR, while DVN-routing performs very closely to the optimal routing based on global information. 

It is worth noting that although AANETs can be formed in many regions where the flight-density is high enough, it may fail in certain regions where the flight-density is low. Future research may integrate low earth orbit satellites into the AANET for supporting truly global coverage.

\bibliographystyle{IEEEtran}

\vspace{-2mm}
\bibliography{ref}
\end{document}